\documentclass[a4paper,11pt]{article}
\usepackage{pos}
\usepackage{physics}
\usepackage{float}
\usepackage{subfig}
\usepackage{tikz}
\usepackage{tabularx}

\usetikzlibrary{quantikz}
\usepackage{tikz-cd}

\title{Lattice QCD results for the topological up-quark mass contribution: too small to rescue the $m_u=0$ solution to the strong CP problem}
\ShortTitle{Lattice QCD results for the topological up-quark mass contribution}

\author[a,b]{Constantia Alexandrou}
\author[b]{Jacob Finkenrath}
\author*[c,d]{Lena Funcke}
\author[e]{Karl Jansen}
\author[f]{Bartosz Kostrzewa}
\author[b,g,h]{Ferenc Pittler}
\author[g,h]{Carsten Urbach}
\affiliation[a]{Department of Physics, University of Cyprus, P.O. Box 20537, 1678 Nicosia, Cyprus}
\affiliation[b]{Computation-Based Science and Technology Research Center, The Cyprus Institute, 20 Kavafi Street, 2121 Nicosia, Cyprus}
\affiliation[c]{Center for Theoretical Physics, Co-Design Center for Quantum Advantage, and NSF AI Institute for Artificial Intelligence and Fundamental Interactions, Massachusetts Institute of Technology, 77 Massachusetts Avenue, Cambridge, MA 02139, USA}
\affiliation[d]{Perimeter Institute for Theoretical Physics, 31 Caroline Street North, Waterloo, ON N2L 2Y5, Canada}
\affiliation[e]{Deutsches Elektronen-Synchrotron DESY, Platanenallee 6, 15738 Zeuthen, Germany}
\affiliation[f]{High Performance Computing \& Analytics Lab, University of Bonn, Friedrich-Hirzebruch-Allee 8, 53115 Bonn, Germany}
\affiliation[g]{Helmholtz Institut f\"ur Strahlen- und Kernphysik, University of Bonn, Nussallee 14-16, 53115 Bonn, Germany}
\affiliation[h]{Bethe Center for Theoretical Physics, University of Bonn, Nussallee 12, 53115 Bonn, Germany}

\emailAdd{c.alexandrou@cyi.ac.cy}  
\emailAdd{j.finkenrath@cyi.ac.cy}
\emailAdd{lfuncke@mit.edu}
\emailAdd{karl.jansen@desy.de}
\emailAdd{kostrzewab@informatik.uni-bonn.de}
\emailAdd{f.pittler@cyi.ac.cy}
\emailAdd{urbach@hiskp.uni-bonn.de}

\abstract{A vanishing Yukawa coupling of the up quark could in principle solve the strong CP problem. To render this solution consistent with current algebra results, the up quark must receive an alternative mass contribution that conserves CP symmetry. Such a contribution could be provided by QCD through non-perturbative topological effects, including instantons. In this talk, we present the first direct lattice computation of this topological mass contribution, using gauge configurations generated by the Extended Twisted Mass collaboration. We use the Iwasaki gauge action, Wilson twisted mass fermions at maximal twist, and dynamical up, down, strange and charm quarks. Our result for the topological mass contribution is an order of magnitude too small to account for the phenomenologically required up-quark mass. This rules out the ``massless'' up-quark solution to the strong CP problem, in accordance with
previous results relying on $\chi$PT fits to lattice data. The talk is based on Ref.~\cite{Alexandrou:2020bkd}, where more details can be found.\\

Preprint number: MIT-CTP/5347 
}

\FullConference{%
 The 38th International Symposium on Lattice Field Theory, LATTICE2021
  26th-30th July, 2021
  Zoom/Gather@Massachusetts Institute of Technology
}

\begin{document}
\maketitle

\section{Introduction}

The strong CP problem is one of the most fundamental open questions of the Standard Model (SM) of particle physics. Its origin is the CP-violating $\theta$-term in the Lagrangian of quantum chromodynamics (QCD) that describes the theory of strong interactions, 
\begin{equation}
\mathcal{L}_{\rm QCD}\supset -\frac{1}{16\pi^2}\, \theta\,G_{\mu\nu}\tilde{G}^{\mu\nu}.
\label{eq:QCDtheta}
\end{equation} 
This $\theta$-term originates from the super-selection sectors of the topologically nontrivial QCD vacuum, which are labeled by an angular parameter $\theta$~\cite{'tHooft1976,'tHooft1976b}. Here, $G^{\mu\nu}$ denotes the gluon field strength and $\tilde{G}_{\mu\nu}=\frac{1}{2}\epsilon_{\mu\nu\rho\sigma} \,G^{\rho\sigma}$ its Hodge dual. The totally antisymmetric Levi-Civita tensor $\epsilon_{\mu\nu\rho\sigma}$ changes sign under parity transformations, which is the reason why the $\theta$-term violates the combined CP symmetry of charge (C) and parity transformations (P).

As $G_{\mu\nu}\tilde{G}^{\mu\nu}$ is a total derivative, one may naively expect that the $\theta$-term disappears after integrating over the Lagrangian in Eq.~\eqref{eq:QCDtheta}. However, the integral $\int d^4 x\, G_{\mu\nu}\tilde{G}^{\mu\nu}\neq 0$ gets nonzero contributions from quantum corrections~\cite{Witten1979a} and instantons~\cite{'tHooft1976}, which are topologically nontrivial field configurations that describe tunneling between the different QCD vacua.
These topologically nontrivial phenomena are known to contribute to the mass of the $\eta'$ meson~\cite{Witten1979b,Veneziano1979}. Thus, the same nonperturbative effects that give a large mass to the $\eta'$ meson are also expected to give rise to strong CP violation via Eq.~\eqref{eq:QCDtheta}.
However, there are strong experimental constraints on CP-violating effects in QCD. These stem from the electric dipole moment $d_n$ of the neutron, which is experimentally excluded down to $|d_n|<3.0 \times 10^{-13}~{\rm e\, fm}$ \cite{Baker2006,Afach:2015sja,Abel:2020}. This translates into a strong upper bound on the angle $|\theta|<1.3 \times 10^{-10}$ due to $d_n\propto m_q\theta$~\cite{Crewther:1979,Crewther1989}. Thus, the $\theta$-parameter needs to be strongly fine-tuned, which is the essence of the strong CP problem. The requirement of strong fine-tuning often hints towards the existence of new physics beyond the SM, in particular towards new symmetries.

One possible solution to the strong CP problem is the Peccei-Quinn (PQ) mechanism \cite{Peccei1977}, which relies on the existence of a chiral $U(1)_{\rm PQ}$ symmetry that is anomalous under the QCD gauge group. 
In this mechanism, the QCD $\theta$-term gets absorbed by rephasing the axion particle $a$, which is the pseudo-Goldstone boson of the spontaneously broken $U(1)_{\rm PQ}$ symmetry~\cite{Weinberg1977,Wilczek1978},
\begin{equation}\label{eq:absorb}
 a \rightarrow a + {\rm const.}\, , \qquad \frac{a}{f_a}\rightarrow \frac{a}{f_a} -\theta.
\end{equation}
Here, $f_a$ is the axion decay constant, and the absorption in Eq.~\eqref{eq:absorb} can happen because the $U(1)_{\rm PQ}$ symmetry is explicitly broken by the Adler-Bell-Jackiw (ABJ) anomaly of QCD~\cite{Adler1969,Bell1969}. 

Within the SM, the simplest realization of an anomalous chiral PQ symmetry $U(1)_{\rm PQ}$ could be achieved if one of the quark flavors, for example the up quark, had no Yukawa coupling to the Higgs doublet. The resulting $U(1)$ symmetry is perturbatively safe and only nonperturbatively broken, which means that it is a true symmetry with regard to 't Hooft's technical naturalness argument~\cite{'tHooft1979}. 
In case of a vanishing up-quark Yukawa coupling to the Higgs doublet, the anomalous chiral PQ symmetry would be an axial $U(1)_{u}$ symmetry acting on the up quark,
\begin{equation}
u \rightarrow e^{i \alpha \gamma_5} u \,,    \qquad \theta\rightarrow\theta+\alpha,
\label{chiralu} 
\end{equation}  
where we combined the left-handed ($u_L$) and right-handed ($u_R$) components of the up quark into a single Dirac fermion $u$. As before, due to the ABJ anomaly, the vacuum $\theta$-angle can be removed by performing the chiral transformation (\ref{chiralu}) and thus becomes unobservable. This scenario is sometimes presented as being fundamentally different from the PQ case, but it actually represents a particular version of the PQ solution: the chiral symmetry (\ref{chiralu}) is spontaneously broken by the QCD up-quark condensate and the role of the axion is played by the $\eta'$ meson (see, e.g., \cite{Dvali2005b,Dvali2013,Dvali2016b}).

This so-called ``massless up-quark solution'' would be the simplest solution to the strong CP problem, as it does not require any new particles or new fundamental energy scales. However, chiral perturbation theory ($\chi$PT) indicates the need for a
nonzero up-quark mass that breaks the chiral symmetry in Eq.~\eqref{chiralu}. In Refs.~\cite{Georgi1981,Kaplan1986, Choi1988,Banks1994}, it was proposed that this nonzero up-quark mass could be generated through the same nonperturbative QCD effects that also contribute to the $\eta'$ mass. This would imply that the
up-quark mass
 in the chiral Lagrangian 
has two different
contributions. First, the perturbative, CP-violating contribution $m_u$ from the Yukawa coupling to the Higgs doublet, which could be easily set to zero by an accidental symmetry~\cite{Leurer1992,Leurer1993,
  Banks1994,Nelson1996,Kaplan1998}. Second, the nonperturbative, CP-conserving contribution $m_{\rm eff}$ from topological effects, such as instantons. Crucially, the latter term does not contribute
to the neutron electric dipole moment and therefore could explain the observed up-quark mass without spoiling the solution to the strong CP problem~\cite{Georgi1981,Kaplan1986,Choi1988,Banks1994,Srednicki:2005wc}.

Due to the nonperturbative nature of the topological mass contribution $m_{\rm eff}$, lattice gauge theory is required to determine its magnitude~\cite{Banks1994,Cohen1999,Dine:2014dga}. Previous lattice computations have focused on the CP-violating mass contribution $m_u$, demonstrating that it is non-vanishing, $m_u(2~{\rm GeV}) = 2.130(41)$~MeV~\cite{Bazavov:2017lyh,Bazavov:2018omf,Aoki:2019cca}. However, these computations relied on fits of the light meson spectrum, and there has been no direct lattice computation of the topological mass contribution itself. 

In a recent paper~\cite{Alexandrou:2020bkd}, we have filled this gap. Based on a theoretical proposal in Refs.~\cite{Cohen1999,Dine:2014dga}, we directly computed the topological up-quark mass contribution by examining the dependence of the pion mass on the dynamical strange-quark mass. This calculation of the topological mass contribution has the advantage of avoiding any fitting procedures. Thus, it provides a complementary analysis of the $m_u=0$ proposal and may finally lay it to rest. Note that both a positive and negative assessment of the $m_u=0$ proposal provides important insights for model building beyond the SM: a positive assessment challenges other proposed solutions, including the
axion~\cite{Peccei1977,Weinberg1977,Wilczek1978} and Nelson-Barr~\cite{Nelson1983,Barr1984} mechanisms, while a negative assessment strengthens the case for these other solutions, which are searched for by several ongoing and planned experiments~(see, e.g., Ref.~\cite{Graham:2015ouw}).

\section{Method}

\begin{figure}[t]
  \centering
  \includegraphics[width=0.48\textwidth]{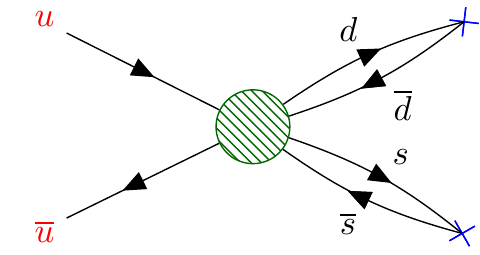} 
  \caption{Topological up-quark mass contribution $m_{\rm eff}=m_dm_s\Lambda_{\rm top}$, where $\Lambda_{\rm top}$ is the characteristic scale of the non-perturbative topological vertex (green circle), such as generated by instantons.}
  \label{fig:top}
\end{figure}

The topological up-quark mass contribution $m_{\rm eff}=m_dm_s/\Lambda_{\rm top}$ (see Fig.~\ref{fig:top}) is known to be proportional to the down-quark mass $m_d$ and the strange-quark mass $m_s$~\citep{'tHooft1986}. The proportionality constant is the inverse characteristic scale $\Lambda_{\rm top}$ of the topological effects, which is unknown and needs to be determined by lattice computations. In our work, we study the variation of the pion mass
\begin{equation}
M_\pi^2 = \beta_1 (m_u+m_d) + \beta_2 m_s (m_u+m_d)+\textnormal{higher orders}\,,
\label{eq:mpisq}
\end{equation}
with respect to the strange-quark mass~\cite{Banks1994}. This variation alters the second term in Eq.~\eqref{eq:mpisq}, which contains both the topological mass contribution (with $1/\Lambda_{\rm top} <\beta_2$) and higher-order corrections in $\chi$PT that are proportional to $m_s$. The first term in Eq.~\eqref{eq:mpisq} stays unaltered and can be used as a reference point in the following way. In order to solve the strong CP problem, the CP-conserving topological mass contribution $m_{\rm eff}=m_dm_s/\Lambda_{\rm top}<\beta_2m_dm_s$ must be large enough to mimic the $m_u$-contribution in the first term of Eq.~\eqref{eq:mpisq}. This gives the following constraint:
\begin{equation}
\frac{\beta_2}{\beta_1}\stackrel{!}{\approx}\frac{m_u}{m_sm_d}\approx 5~\mathrm{GeV}^{-1}
\label{eq:solution}
\end{equation}
at renormalization scale $\bar\mu =
2~\mathrm{GeV}$ in the $\overline{\textrm{MS}}$ scheme~\cite{Georgi1981,Kaplan1986,Choi1988,Banks1994, Cohen1999,Dine:2014dga}.

In our lattice computation~\cite{Alexandrou:2020bkd}, we assume equal and fixed masses of the lightest quarks, $m_u=m_d\equiv m_\ell$, and we vary the strange-quark mass $m_{s,i}$, which determines $\beta_2/\beta_1$ via~\cite{Cohen1999,Dine:2014dga}
\begin{equation}
 \frac{\beta_2}{\beta_1}\ = \frac{M_{\pi,1}^2 - M_{\pi,2}^2}{m_{s,1}
   M^2_{\pi,2} -m_{s,2} M^2_{\pi,1}}\, \Bigg|_{M_{\pi,i}\to 0}.
 \label{eq:beta}
\end{equation}
Here, $M_{\pi,i}=M_\pi(m_{s,i})$ is the charged pion mass defined in Eq.~\eqref{eq:mpisq}. Note that $\beta_2/\beta_1$ in Eq.~\eqref{eq:beta} only becomes exact after the chiral extrapolation $M_{\pi,i}\to 0$, which ensures the cancellation of higher-order corrections in Eq.~\eqref{eq:mpisq}. As the ratio $\beta_2/\beta_1$ is independent of $m_\ell$, we can reliably compute this ratio using pion masses $M_\pi(m_\ell,m_{s})$ with $m_\ell$ larger than its physical value.

\begin{table}
  \centering
  \begin{tabular*}{\textwidth}{@{\extracolsep{\fill}}lcccccccc}
    \hline\hline \\[-2.0ex]
    Ensemble  & $a\mu_\ell$ & $a\mu_\sigma$ 
    & $a\mu_\delta$ & $aM_\pi$ & $M_\pi~[\mathrm{MeV}]$ & 
    $m_s~[\mathrm{MeV}]$\\
    \hline\hline  \\[-2.0ex]
      A60  & $0.006$ & $0.15$ & $0.190$ & 0.17308(32) & $386(16)$ & $98(4)$\\
    A60s &           &        & $0.197$ & 0.17361(31)&  $387(16)$ & $79(4)$\\
    A80  &       $0.008$ &        & $0.190$ & 0.19922(30)&  $444(18)$ & $98(4)$\\
    A80s &            &        & $0.197$ & 0.19895(42)&  $443(18)$ & $79(4)$\\
    A100 &      $0.010$ &        & $0.190$ & 0.22161(35)& $494(20)$ & $100(4)$\\
    A100s&          &        & $0.197$ & 0.22207(27)& $495(20)$ & $79(4)$\\
    \hline  \\[-2.0ex]
    cA211.30.32 & $0.003$ & $0.1408$ & $0.1521$
    & $0.12530(14)$ & $276(3)$ & $99(2)$\\
    cA211.30.32l & & $0.1402$ & $0.1529$
    & $0.12509(16)$& $275(3)$ & $\phantom{0}94(2)$\\
    cA211.30.32h & & $0.1414$ & $0.1513$
    & $0.12537(14)$& $276(3)$ & $104(2)$\\
    \hline\hline
  \end{tabular*}
  \caption{Parameters of the two different types of ensembles used in our computations. All dimensionful quantities are quoted
  in units of the lattice spacing $a$, unless denoted otherwise. Here, $\mu_\ell$ is the bare mass of the light quarks. The parameters $\mu_\sigma$ and $\mu_\delta$ determine the renormalized
strange and charm quark masses via  $m_s = (\mu_\sigma/Z_P) - (\mu_\delta/Z_S)$ and $m_c
 = (\mu_\sigma/Z_P)   + (\mu_\delta/Z_S) $, where $Z_P$ and $Z_S$ are the pseudoscalar and scalar renormalization
functions, respectively. The pion mass $M_\pi$ is given both in units of $a$ and in physical units. The strange-quark mass $m_s$ is given at
    $2~\mathrm{GeV}$ in the $\overline{\mathrm{MS}}$ scheme. Table adapted from Ref.~\cite{Alexandrou:2020bkd}.}
  \label{tab:parameters}
\end{table}

\section{Lattice computation}

In our computation of the topological up-quark mass contribution~\cite{Alexandrou:2020bkd}, we used dynamical up, down, strange, and charm quark flavors with degenerate masses of the lightest quarks. Our gauge
configurations were generated by the Extended Twisted Mass (ETM) Collaboration, using the Iwasaki improved gauge action~\cite{Iwasaki:1985we} and Wilson twisted mass fermions, $\psi(x)\to\exp(-i\omega\gamma_5\tau^3/s)\psi(x)$, $m_\psi\to\exp(i\omega\gamma_5\tau^3)m_\psi$, at maximal twist, $\omega=\pi/2$~\cite{Frezzotti:2000nk,Frezzotti:2003xj}. 
In Table~\ref{tab:parameters}, we list all the ensembles, pion masses, and quark masses that we used in our study (for more details, see the supplemental material of~\cite{Alexandrou:2020bkd}).
In particular, we used three pairs of ensembles (AX and AXs) with a lattice spacing value of $a = 0.0885(36)\ \mathrm{fm}$~\cite{Carrasco:2014cwa} and without a clover term in the action~\cite{Baron:2010bv}, as well as one ensemble (cA211.30.32) with $a=0.0896(10)\ \mathrm{fm}$~\cite{Alexandrou:2020bkd} and a
clover term~\cite{Alexandrou:2018egz}.
All seven ensembles stem from simulations of several thousand trajectories. This, together with their relatively coarse lattice spacing, ensures that topological sectors are well sampled.

Each of the three pairs of ensembles (AX and its corresponding AXs) has different values for $m_s$ and $m_c$ but otherwise identical parameters. The difference between the three pairs with $X=60$, $80$, and $100$ is the different values for $m_\ell$ corresponding
to $M_\pi=386~\mathrm{MeV}$, $M_\pi=444~\mathrm{MeV}$ and
$M_\pi=494~\mathrm{MeV}$, respectively. We use these three different pairs of ensembles with different values for $m_s$ and $M_\pi$ to directly compute $\beta_2/\beta_1$ from Eq.~\eqref{eq:beta}. 

\begin{figure}[t]
  \centering
  \includegraphics[width=0.7\textwidth]{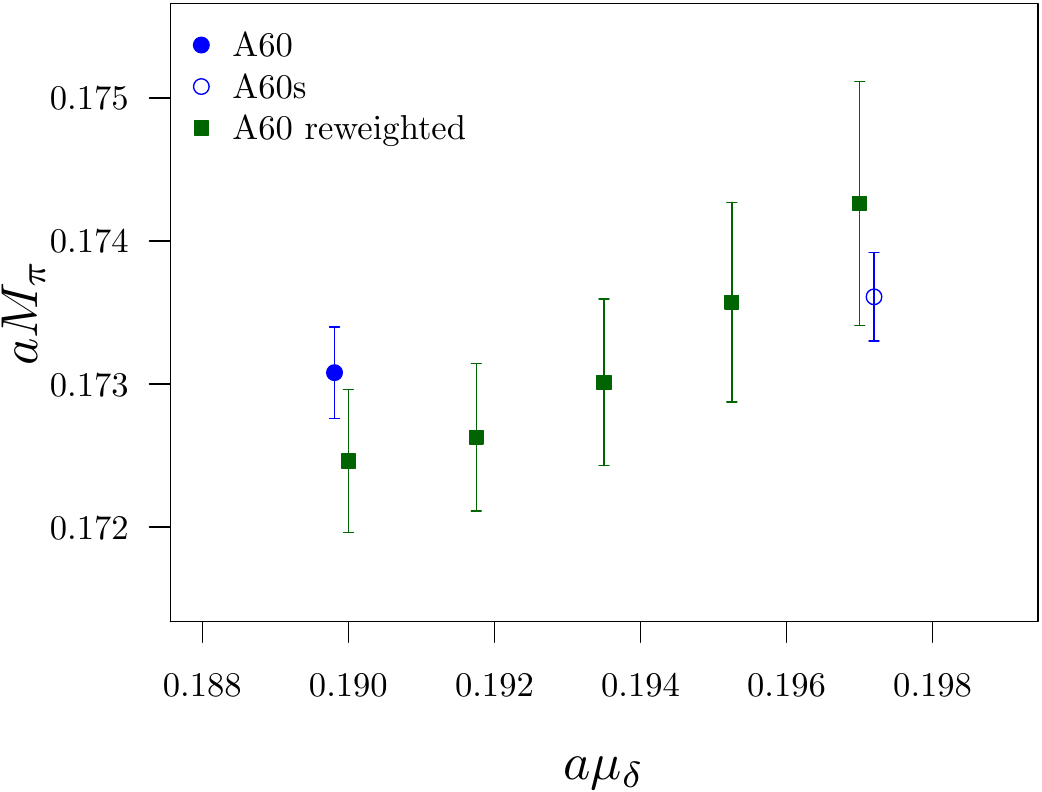} 
  \caption{Pion mass $M_\pi$ as a function of $a\mu_\delta$. Original results for the A60 (filled blue circle) and A60s (open blue circle) ensembles, compared to reweighting results (green boxes). Figure taken from Ref.~\cite{Alexandrou:2020bkd}.}
  \label{fig:reweight}
\end{figure}

For the ensemble cA211.30.32, the values for $m_s$ and $m_c$ are similar to the ones for the AX(s) ensembles, but the pion mass is much smaller, $M_\pi=270~\mathrm{MeV}$, thus closer to the physical value. The most crucial difference to the AX(s) ensembles is that the cA211.30.32 ensemble only has a single strange-quark mass value, which prevents a direct computation of $\beta_2/\beta_1$ from Eq.~\eqref{eq:beta}. Therefore, we compute the different $M_\pi(m_{s,i})$ and $m_{s,i}$ required in Eq.~\eqref{eq:beta} through reweighting. Here, we call cA211.30.32h (cA211.30.32l) the reweighted ensemble with a 5\% higher (lower) value for $m_s$ than the original ensemble cA211.30.32. We also perform a cross-check of the reweighting procedure using the AX(s) ensembles, as shown in Fig.~\ref{fig:reweight}. Here, we split the reweighting factor in several steps, which allows us to compute $M_\pi$ for three intermediate steps (green boxes) between the original values of $m_s$ for the A60 (filled blue circle) and A60s (open blue circle) ensembles. 

Thus, using these two different types of ensembles, AX(s) and cA211.30.32, allows us to both test the reweighting procedure and study the $M_\pi$-dependence of Eq.~\eqref{eq:beta}, which will later enable a reliable chiral extrapolation to determine $\beta_2/\beta_1$ for $M_\pi\to 0$.

\section{Results}

Our results for the ratio $\beta_2/\beta_1$ are shown in Table~\ref{tab:beta12}, which we obtained using Eq.~(\ref{eq:beta}) with the input values for $M_\pi$ and $m_s$ given in lattice units, see Table~\ref{tab:parameters}. We denote with cA211.30.32(h) the results obtained with input values from the original ensemble cA211.30.32 and the reweighted ensemble
cA211.30.32h. Similarly, cA211.30.32(l) corresponds to the ensembles cA211.30.32 and cA211.30.32l, while cA211.30.32(h,l) corresponds to the ensembles cA211.30.32h and cA211.30.32l. Our results for $\beta_1$ are strictly positive, which is expected as this coefficient is proportional to the chiral condensate \cite{Novikov1981},
\begin{equation}
\frac{d\beta_1}{dm_s}\propto\frac{d}{dm_s}\langle\bar{\psi}\psi\rangle\propto\int_{{\rm lim}\, k\to 0}dx e^{ikx}\langle\bar{\psi}\psi(x),\bar{s}s(0)\rangle.
\end{equation}
Note that our results for $\beta_2$ are compatible with zero, and $\beta_2/\beta_1$ is zero at the $1.5\sigma$ level.

\begin{figure}[t]
  \centering
  \includegraphics[width=0.7\textwidth]{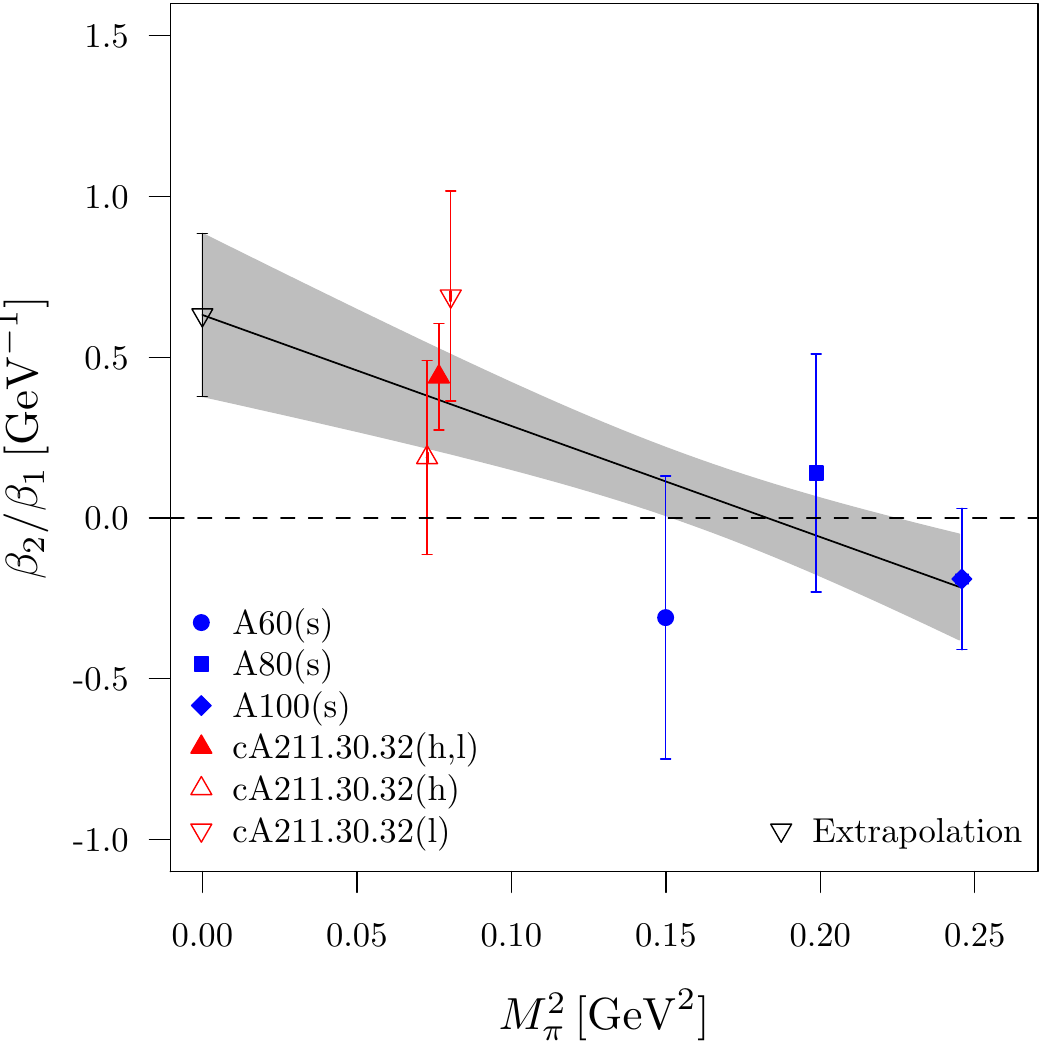} 
  \caption{Results for the ratio $\beta_2/\beta_1$ as a function of $M_\pi^2$ in physical units. The ensembles are AX(s) without a clover term (blue) and cA211.30.32 with a clover term (red). For better legibility, the red data points are displaced horizontally. The solid line (black) with the
    $1\sigma$ error band (grey) denotes the linear chiral extrapolation in $M_\pi^2$, see Eqs.~\eqref{eq:beta} and \eqref{chPT}. Figure taken from Ref.~\cite{Alexandrou:2020bkd}.}
  \label{fig:beta21}
\end{figure}

In Fig.~\ref{fig:beta21}, we plot our results from Table~\ref{tab:beta12} as a function of $M_\pi^2$. The blue (red) data points are the results for the AX(s) ensembles (the cA211.30.32 ensemble). The black line is the chiral extrapolation that eliminates higher-order corrections to $\beta_2/\beta_1$, as explained below Eq.~\eqref{eq:beta}. We choose a linear extrapolation due to the $\chi$PT prediction of~\cite{Aoki:2019cca} 
\begin{equation}
  \frac{\beta_2}{\beta_1} \approx \frac{\alpha_2}{\alpha_1 +(\alpha_3
    /\alpha_1)M_\pi^2} \approx \frac{\alpha_2}{\alpha_1} -
  \frac{\alpha_2\alpha_3}{\alpha_1^3}M_\pi^2, \label{chPT}
\end{equation}
modulo logarithmic corrections. Here, the coefficients $\alpha_{1,2,3}$ are combinations of low-energy constants with
$\alpha_1\gg (\alpha_3/\alpha_1)M_\pi^2$, $M_\pi^2=\alpha_1 m_\ell
+\mathcal{O}(\alpha_{2,3})$, and $\mathcal{O}(\alpha_{2,3})/ (\alpha_1
m_\ell)\approx 0.1$.
For the linear fit, we use all three data points from the AX(s) ensemble but only one data point from the cA211.30.32 ensemble (see the filled symbols in Fig.~\ref{fig:beta21}), because the three data points for the ensemble cA211.30.32
are strongly correlated. The fit has a $p$-value of $0.2$ and yields a chirally extrapolated result of $\beta_2/\beta_1 (M_\pi\to 0)= 0.63(25)~\mathrm{GeV}^{-1}$. 
Taking into account the $1\sigma$ statistical uncertainty, this result excludes the
value of $\beta_2/\beta_1\approx 5~\mathrm{GeV}^{-1}$ that is required to solve the strong CP problem by more than $10\sigma$, see Eq.~\eqref{eq:solution}.

\begin{table}
  \centering
  \begin{tabular*}{\textwidth}{@{\extracolsep{\fill}}lccc}
    \hline\hline  \\[-2.0ex]
    Ensemble & $\beta_2~[\mathrm{GeV^2}]$ &
    $\beta_1~[\mathrm{GeV^3}]$ &
    $\beta_2/\beta_1~[\mathrm{GeV}^{-1}]$\\
    \hline\hline  \\[-2.0ex]
    A60(s)  & $-0.0009(08)$ & $\phantom{-}0.0029(4)$ & $-0.32(26)$\\
    A80(s)  & $\phantom{-}0.0005(10)$ & $\phantom{-}0.0036(4)$ & $\phantom{-}0.15(30)$\\
    A100(s) & $-0.0010(10)$ & $\phantom{-}0.0053(6)$ & $-0.19(19)$\\
    \hline \\[-2.0ex]
    cA211.30.32(h)   & $0.00007(11)$ & $0.00039(5)$ & $\phantom{-}0.18(30)$ \\
    cA211.30.32(l)   & $0.00026(11)$ & $0.00037(5)$ & $\phantom{-}0.69(33)$ \\
    cA211.30.32(h,l) & $0.00033(12)$ & $0.00076(5)$ & $\phantom{-}0.43(16)$ \\
    \hline\hline
  \end{tabular*}
  \caption{Results for the coefficients $\beta_2$ and $\beta_1$ defined in Eq.~\eqref{eq:mpisq} and their ratio $\beta_2/\beta_1$. All results are obtained from Eq.~(\ref{eq:beta}) and given in physical units at
    $\bar\mu=2~\mathrm{GeV}$ in the 
    $\overline{\mathrm{MS}}$ scheme. Table taken from Ref.~\cite{Alexandrou:2020bkd}.}
  \label{tab:beta12}
\end{table}

To estimate the systematic uncertainties, the discretization errors for $\beta_2/\beta_1$ can be obtained by comparing our lattice results for $m_s$ and $M_\pi$ to the known continuum
extrapolation values for the AX ensembles~\cite{Carrasco:2014cwa}. The resulting discretization errors are of the order of $5-10\%$, which implies errors of maximally $10\%$
for $\beta_2$, $15\%$ for $\beta_1$, and $20\%$  for $\beta_2/\beta_1$, because most lattice artifacts cancel in the differences in Eq.~\eqref{eq:beta}. For the ensemble cA211.30.32, the lattice artifacts are even further reduced due to the inclusion of the clover term~\cite{Alexandrou:2020bkd}.
There are no finite-size effects for $m_s$, and the finite-size corrections for $M_\pi$ are equal for $M_{\pi,1}^2$ and $M_{\pi,2}^2$, thus canceling for the ratio $\beta_2/\beta_1$.

In total, our chirally extrapolated value for $\beta_2/\beta_1$ has a $1\sigma$ statistical error and a conservatively estimated $20\%$ systematic error from lattice artifacts. Thus, we arrive at the following result~\cite{Alexandrou:2020bkd}:
\begin{equation}
  \begin{split}
   \frac{\beta_2}{\beta_1}\ =\ 0.63(25)_\mathrm{stat}(14)_\mathrm{sys}~\mathrm{GeV}^{-1} \ =\ 0.63(39)~\mathrm{GeV}^{-1}\\
  \end{split}
  \label{eq:bound}
\end{equation}
at $\bar\mu=2~\mathrm{GeV}$ in the $\overline{\textrm{MS}}$
scheme. We note that $\beta_2/\beta_1$ receives logarithmic corrections~\cite{Novikov1981,Gasser1985}, which are of the same order as our result in Eq.~\eqref{eq:bound}; this renders the topological contribution to $\beta_2/\beta_1$ even smaller. We also note that a constant extrapolation in $M_\pi^2$ would have been equally well compatible with our data and would have given a substantially smaller result for $\beta_2/\beta_1$. Thus, our result can be considered as a conservative upper bound for $\beta_2/\beta_1$.

\section{Conclusion}

In our work~\cite{Alexandrou:2020bkd}, we have provided the first direct lattice computation of the topological up-quark mass contribution, by studying the dependence of the pion mass on the dynamical strange-quark mass. Using Wilson twisted mass fermions at maximal twist, the Iwasaki gauge action, and gauge configurations generated by the ETM Collaboration, we determined an upper bound for the strength of the topological mass contribution, $\beta_2/\beta_1<1.02~{\rm GeV}^{-1}$, see Eq.~(\ref{eq:bound}). Our systematic error estimates are highly conservative, and our result is significantly lower than the value of $\beta_2/\beta_1\approx 5~{\rm GeV}^{-1}$ required by the massless up-quark solution to the strong CP problem. Thus, our work excludes the massless up-quark solution, in agreement with previous results using direct $\chi$PT fits of the light meson spectrum. These findings strengthen the case for alternative solutions to the strong CP problem, including the axion solution, which are highly sought after experimentally.

\begin{acknowledgments}
We thank all members of the ETM Collaboration for the most enjoyable
  collaboration. We also thank J.~Gasser, D.~Kaplan, T.~Banks, Y.~Nir, and N.~Seiberg, and U.-G.~Meißner for useful comments and discussions.
  We kindly thank F.~Manigrasso and K.~Hadjiyiannakou for providing 
  the necessary correlators for computing the nucleon mass on the ensemble
  cA211.30.32. 
  The authors gratefully acknowledge the Gauss Centre for Supercomputing (GCS)
  e.V. (www.gauss-centre.eu) for funding this project by providing
  computing time on the GCS supercomputer JUQUEEN~\cite{juqueen} and the
  John von Neumann Institute for Computing (NIC) for computing time
  provided on the supercomputers JURECA~\cite{jureca} and JUWELS at Jülich
  Supercomputing Centre (JSC) under the projects hch02, ecy00, and hbn28.
  The project used resources of the SuperMUC at the Leibniz Supercomputing
  Centre under the Gauss Centre for Supercomputing e.V.
  project pr74yo.
  The cA211.30.32 ensemble was generated on the Marconi-KNL supercomputer
at CINECA within PRACE project Pra13-3304.
  This project was funded in part by the DFG as a project in the
  Sino-German CRC110 (TRR110) and by the PRACE Fifth and Sixth Implementation Phase 
  (PRACE-5IP and PRACE-6IP) program of the European Commission under
  Grant Agreements No.\ 730913 and No.\ 823767. Research at Perimeter Institute is supported in part by the Government of Canada through the Department of Innovation, Science and Industry Canada and by the Province of Ontario through the Ministry of Colleges and Universities. L.F.\ is partially supported by the U.S.\ Department of Energy, Office of Science, National Quantum Information Science Research Centers, Co-design Center for Quantum Advantage (C$^2$QA) under contract number DE-SC0012704, by the DOE QuantiSED Consortium under subcontract number 675352, by the National Science Foundation under Cooperative Agreement PHY-2019786 (The NSF AI Institute for Artificial Intelligence and Fundamental Interactions, http://iaifi.org/), and by the U.S.\ Department of Energy, Office of Science, Office of Nuclear Physics under grant contract numbers DE-SC0011090 and DE-SC0021006.
   J.F. is financially supported by H2020 project PRACE 6-IP (grant agreement No 82376) and the EuroCC project funded by the Deputy Ministry of Research, Innovation and Digital Policy and the Cyprus Research and Innovation Foundation and
the European High-Performance Computing Joint Undertaking (JU) under grant agreement No 951732. The JU receives
support from the European Union’s Horizon 2020 research and innovation programme. F.P.\ acknowledges financial support from the Cyprus Research and Innovation Foundation under project ''NextQCD``, contract number EXCELLENCE/0918/0129.
  The open source software packages tmLQCD~\cite{Jansen:2009xp,Deuzeman:2013xaa,Abdel-Rehim:2013wba},
  Lemon~\cite{Deuzeman:2011wz}, DDalphaAMG~\cite{Frommer:2013fsa,Alexandrou:2016izb,Alexandrou:2018wiv},
  QUDA~\cite{Clark:2009wm,Babich:2011np,Clark:2016rdz}, and R~\cite{R:2005} have been used.\looseness=-1
\end{acknowledgments}

\bibliographystyle{JHEP}
\bibliography{bibliography}

\end{document}